\documentclass{webofc}
\usepackage[varg]{txfonts}   
\newcommand{\AGeVc}{$A$~GeV/$c$}
\newcommand{\AGeV}{$A$~GeV}
\newcommand{\A}{$A$}
\newcommand{\SHINE}{NA61/SHINE}
\begin{document}
	\pagenumbering{arabic}
	\title{Effects of the detector non-uniformity in pion directed flow measurement relative to the spectator plane by the NA49 experiment at the CERN SPS}
	%
	%
	
	\author{\firstname{Oleg} \lastname{Golosov}\inst{1}\fnsep\thanks{\email{oleg.golosov@gmail.com}} \and
		\firstname{Evgeny} \lastname{Kashirin}\inst{1} \and
		\firstname{Viktor} \lastname{Klochkov}\inst{2,3} \and
		\firstname{Ilya} \lastname{Selyuzhenkov}\inst{1,2} \\
		\vspace{-0.2cm}\\
		\centering{for the NA49 Collaboration}
	}
	
	\institute{National Research Nuclear University (Moscow Engineering Physics Institute) Moscow, Russia 
		\and
		GSI Helmholtzzentrum f\"ur Schwerionenforschung, Darmstadt, Germany 
		\and
		Goethe-University Frankfurt, Frankfurt, Germany
	}
	
	\abstract{
		Results are presented for directed flow $v_1$ of negatively charged pions measured relative to the spectator plane in Pb+Pb collisions at the beam energy 40\AGeV~recorded by the NA49 experiment at CERN. The measurements were performed as a function of rapidity and transverse momentum in two classes of collision centrality.
		The projectile spectator symmetry plane is estimated using the transverse segmentation of the forward hadron calorimeters.
		Analysis details related to event selection and particle identification are provided along with a description of the procedure used to apply corrections for the limited detection efficiency. Additionally a systematic study of effects due to detector non-uniformity on pion directed flow measurement is presented.
	}
	\maketitle
	\section{The NA49 experiment and data sample}
	\label{data}
	The NA49 experiment at the CERN SPS is the predecessor of the currently operating fixed target \SHINE~experiment which has recently extended its program with a Pb-ion beam momentum scan at 13\A, 30\A~and 150\AGeVc.
	NA49 collected data for Pb+Pb collisions at beam energies of 20\A, 30\A, 40\A, 80\A { and 158\AGeV}~\cite{Alt:2003ab}.
	The NA49 and \SHINE~data complement the measurement of flow harmonics available from the Beam Energy Scan (BES) program of STAR at  RHIC~\cite{Adamczyk:2014ipa} and extend the measurement towards forward rapidity up to the region where projectile spectators appear. 
	The new results are also relevant for studies in the few GeV collision energy range, in particular for the preparation of the Compressed Baryonic Matter (CBM) heavy-ion experiment at the future FAIR facility in Darmstadt.
	
	A description of the NA49 experiment layout and its trigger system can be found in~\cite{AFANASIEV1999210}.
	It had four large volume time-projection chambers (TPC) - two vertex TPCs (VTPC-1 and VTPC-2) positioned inside the magnets and two main TPCs (MTPC-L and MTPC-R) which are used
	for momentum reconstruction and particle identification via measurement of the specific energy loss ($dE/dx$).
	
	The ring calorimeter (RCAL) \cite{DEMARZO1983405} was positioned at 18 meters from the target and its transverse granularity of 10 rings in radial and 24 modules in azimuthal direction is used to estimate the spectator plane resolution of the veto calorimeter.
	The veto (VCAL) \cite{AFANASIEV1999210} calorimeter was installed 20 meters downstream of the target behind a collimator and has a 2x2 transverse module granularity. The opening of the collimator is adjusted such that beam particles, projectile and spectator fragments can reach the VCAL.
	\begin{figure}[h]
		\centering
		\includegraphics[width=0.99\textwidth]{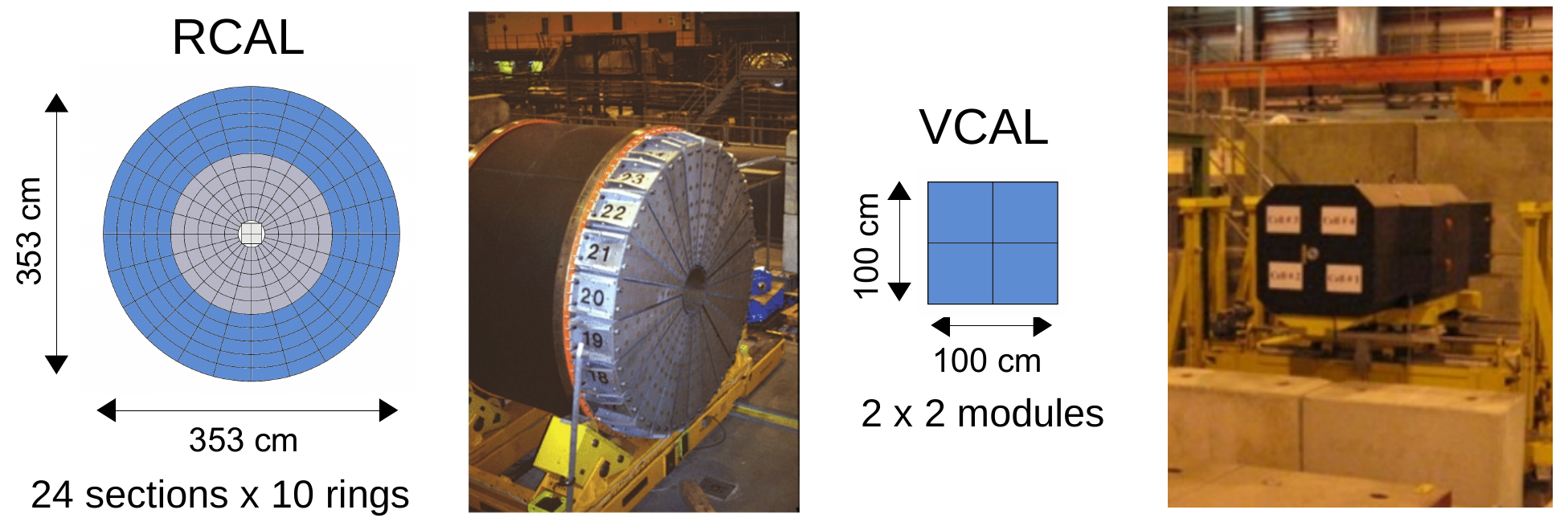} 
		\caption{Schematic view and images of ring calorimeter (RCAL - left) and veto calorimeter (VCAL - right).}	
		\label{Fig:calorimeters}
	\end{figure}

	The present analysis used a sample of Pb+Pb collisions at 40\AGeV~recorded in the year 2000 with minimum bias, midcentral and central triggers which are fully efficient for centrality classes 0-60\%, 0-40\% and 0-12.5\% respectively.
	More details about centrality determination are presented in~\cite{Kashirin:2018syo}.
	Events with fitted vertex position within and close to the target region and containing at least 10 tracks associated with the primary vertex were selected.
	20k events were selected for minimum bias, 340k for midcentral and 440k for central triggers.
	
	Only tracks with at least 20 points in the VTPCs and at least 30 points in all TPCs were considered.
	To exclude split tracks the number of hits associated with a track was required to be more than 55\% of total hits possible for a given track.
	Primary tracks were selected using the distance of closest approach (DCA) to the fitted primary vertex in the plane transverse to the beam. Only tracks with DCA less than $2 ~cm$ in the magnetic field bending ($x$) direction and less than $1 ~cm$ in the perpendicular ($y$) direction were considered. 
	\begin{figure}[htb]
		\noindent
		\includegraphics[width=0.495\textwidth]{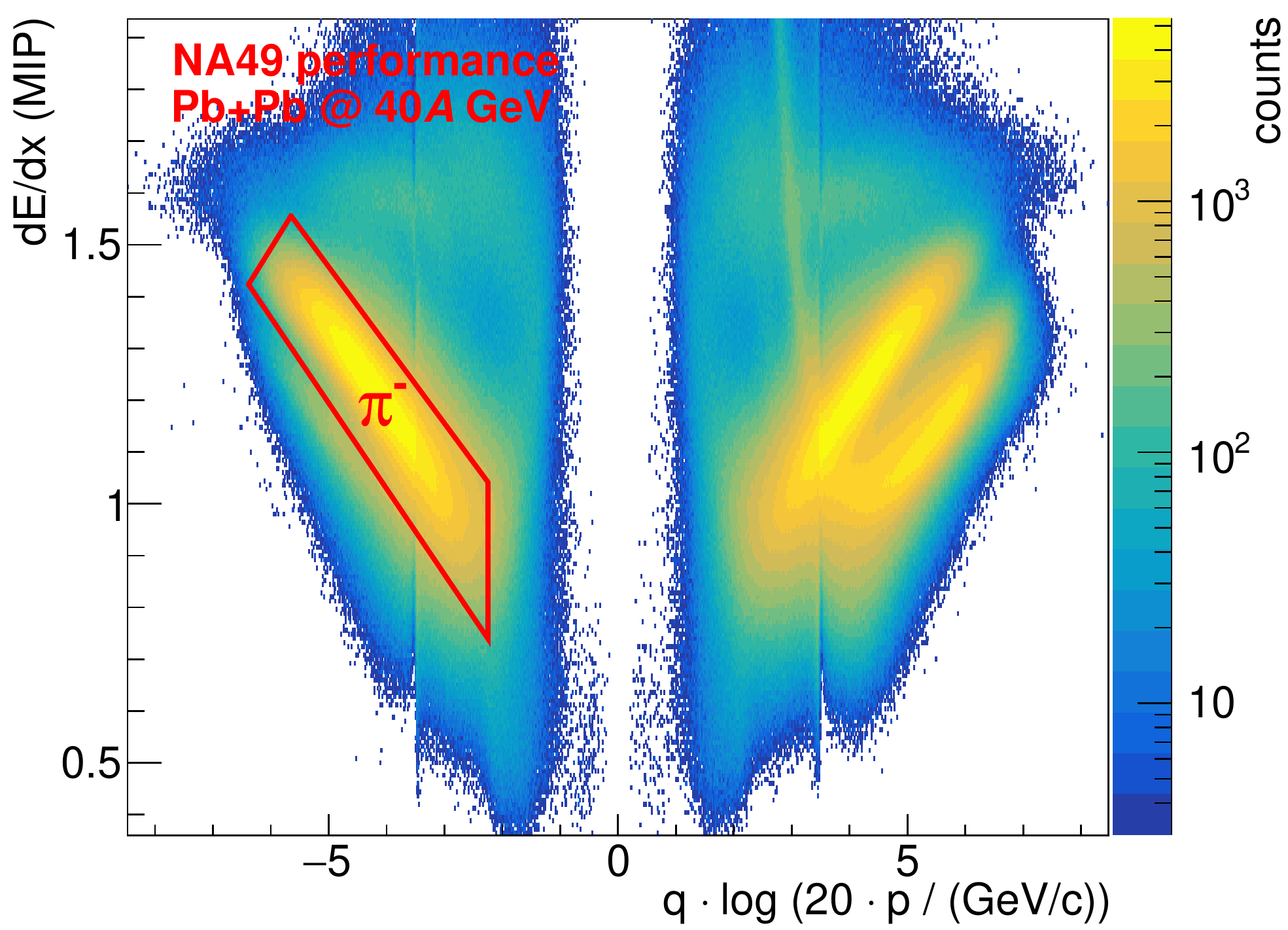}
		\includegraphics[width=0.495\textwidth]{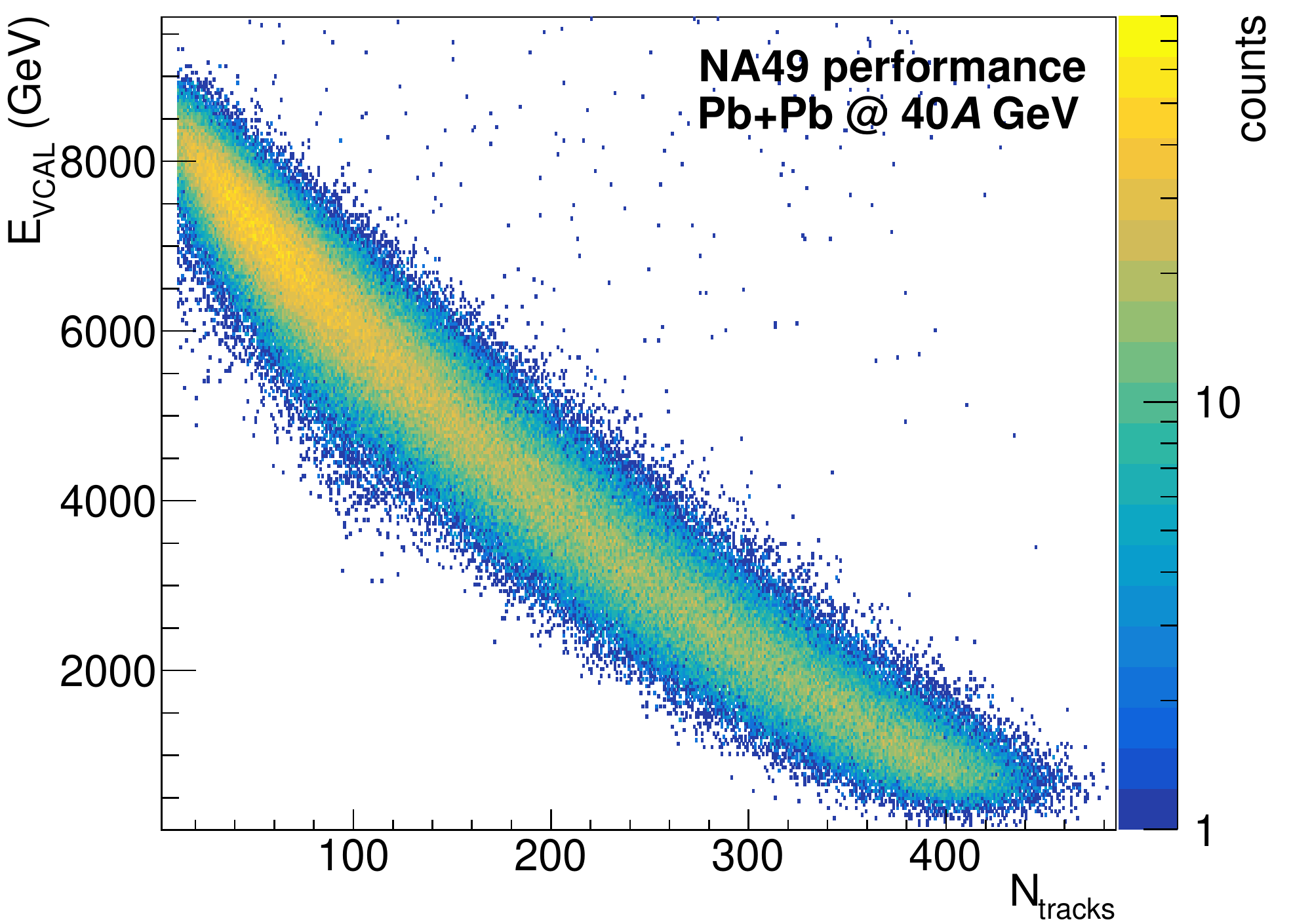}
		\caption{
			Left: track $dE/dx$ as a function of momentum for different charges. Region used for negatively charged pion selection is indicated by the red box.
			Right: Anticorrelation between VCAL energy and number of primary tracks in the TPCs.
		}
		\label{Fig:centrality}
	\end{figure}
	Negatively charged pions were selected using $dE/dx$ measurements from all TPCs (Fig. \ref{Fig:centrality}, left). 
	
	The energy deposited in the VCAL modules comes mostly from projectile spectators and is used for collision centrality and symmetry plane estimation.
	Figure \ref{Fig:centrality} (right) shows the expected anticorrelation between projectile spectator energy measured with VCAL and the number of tracks reconstructed with the TPCs.

	\section{Flow analysis details}
	\label{analysis}
	
	Directed flow $v_1$ is measured from the correlation of two-dimensional flow vectors $\mathbf{q_1}$ and $\mathbf{Q_1}$. The vector $\mathbf{q_1}$ is calculated event-by-event from the azimuthal angles $\phi_i$ of selected pions:
	\begin{equation}
	\label{Eq:q1_TPC}
	\mathbf{q_1} = \frac{1}{W} \sum_{i=1}^{M} w_i \mathbf{u_{1,i}},~~~~~~~~~~~W = \sum_{i=1}^{M} w_i,
	\end{equation}
	where $M$ is the number of selected pions in the event, $i$ the pion track index, $w_i$ a weight, accounting for detection efficiency in the given kinematic bin (see Sec. \ref{corrections}), and $\mathbf{u_{1,i}} = ( \cos{\phi_i}, \sin{\phi_i})$.
	The spectator symmetry plane is estimated with the $\mathbf{Q_1}$ direction determined from the azimuthal asymmetry of the energy deposition in VCAL:
	\begin{equation}
	\mathbf{Q_1^{\rm VCAL}} = \frac{1}{E_{\rm VCAL}} \sum_{i=1}^{4} E_i \mathbf{n_i},~~~~~~~~~E_{\rm VCAL} = \sum_{i=1}^{4} E_i,
	\label{Eq:Q_VCAL}
	\end{equation}
	where the unit vector $\mathbf{n_i}$ points in the direction of the center of i-th VCAL module and $E_i$ is the energy deposited in it.
	An equation similar to Eq.~(\ref{Eq:Q_VCAL}) is used to calculate $\mathbf{Q_1^{\rm RCAL1}}$ and $\mathbf{Q_1^{\rm RCAL2}}$. For this rings of RCAL modules are divided into two subgroups: 5 inner (RCAL1) and 5 outer (RCAL2) rings.
	
	Independent estimates of directed flow $v_1$ are obtained using the scalar product method:
	\begin{equation}
	v_1^\alpha \lbrace A|B,C \rbrace = \frac{2 \langle q_{1, \alpha} Q^A_{1,\alpha} \rangle}{R^A_{1,\alpha} \lbrace B, C \rbrace},
	\end{equation}
	where $\alpha = x, y$ are $\mathbf{q_1}$ and $\mathbf{Q_1}$ components and $A$=VCAL, $B$=RCAL1, and $C$=RCAL2. Symmetry plane resolution correction factors $R^A_{1,\alpha} \lbrace B, C \rbrace$ were calculated with the three-subevent method:
	
	\begin{equation}
	\label{resolution}
	R^A_{1,\alpha} \lbrace B, C \rbrace = \sqrt{2 \frac{
			\langle Q^A_{1, \alpha} Q^B_{1, \alpha} \rangle \langle Q^A_{1, \alpha} Q^C_{1, \alpha} \rangle 
		}{
		\langle Q^B_{1, \alpha} Q^C_{1, \alpha} \rangle
	}}.
	\end{equation}
	
	\section{Acceptance, efficiency and systematic checks}
	\label{corrections}
\begin{figure}[htb]
		\noindent
		\includegraphics[width=0.49\textwidth]{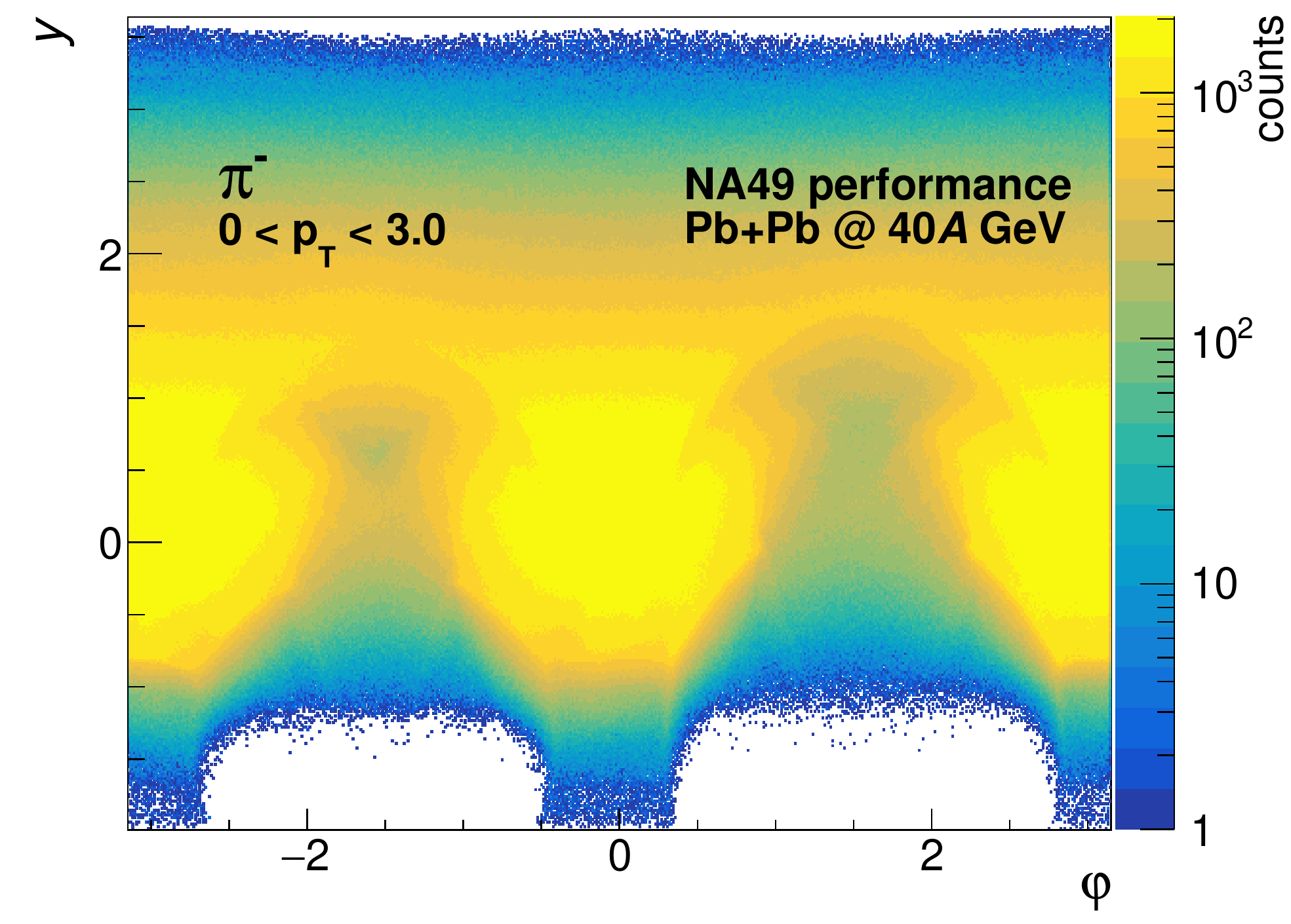}
		\includegraphics[width=0.49\textwidth]{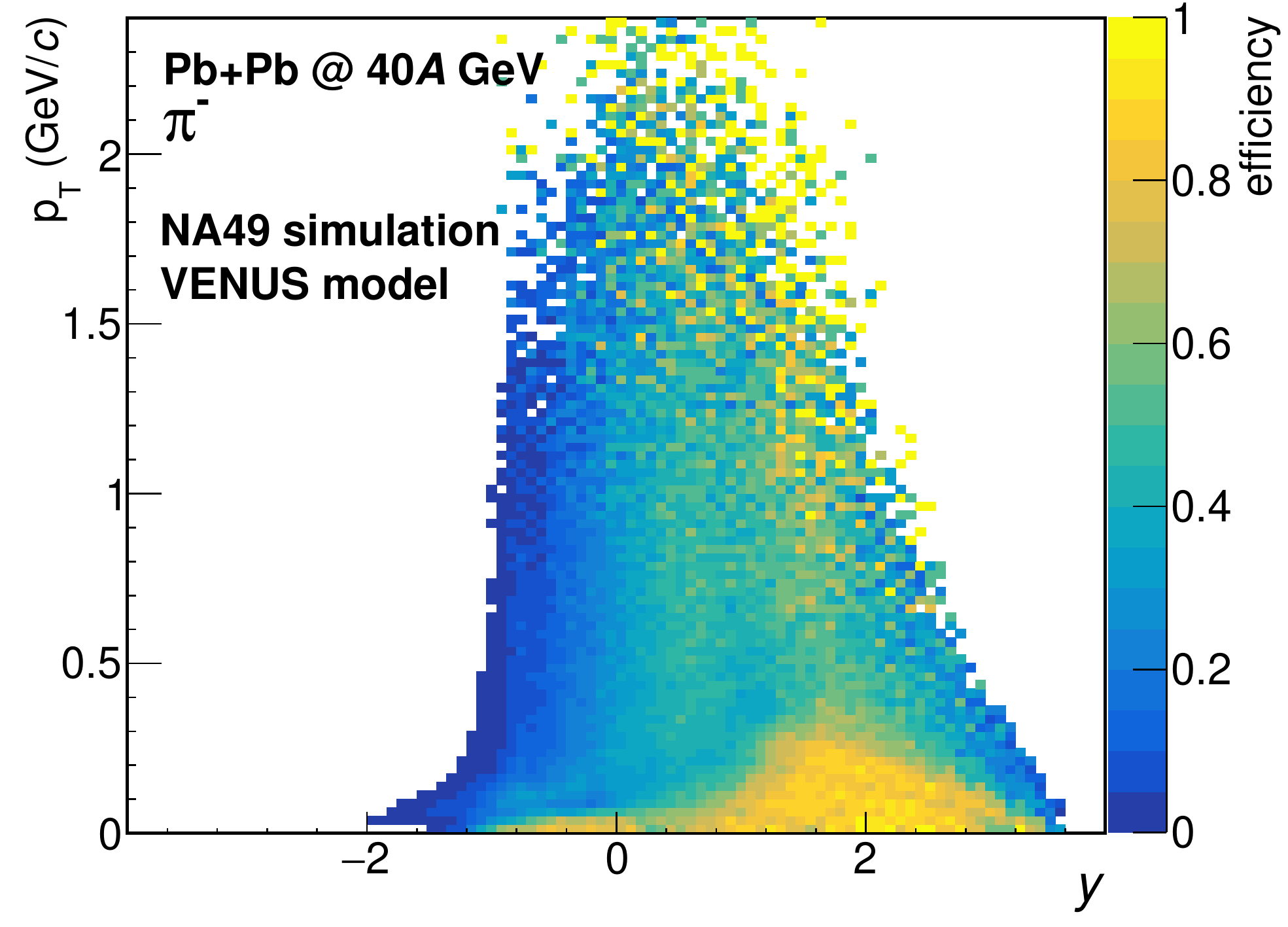}
		\caption{
			Left: Distribution of azimuthal angles and rapidity for negatively charged pions.
			Right: Efficiency as a function of transverse momentum and rapidity for negatively charged pions.}
		\label{Fig:corrections}
	\end{figure}

	Due to the rectangular shape of the NA49 TPCs their azimuthal acceptance in the transverse plane is highly non-uniform and introduces a substantial bias in flow measurements. Non-uniformity may vary with event and track properties (see Fig.~\ref{Fig:corrections}, left). This effect was accounted for using a data-driven procedure described in Ref.~\cite{Selyuzhenkov:2007zi} and implemented in the QnCorrections framework~\cite{QnGithub:2015,Gonzalez:2016GSI}. Recentering, twist and rescaling corrections were applied as a function of time and collision centrality for \textbf{Q} and \textbf{q}-vectors and for \textbf{q}-vectors additionally as a function of track transverse momentum and rapidity. 
	
	Effects of transverse momentum and rapidity dependent detection efficiency on the measurement of $v_1$ were corrected using a Monte-Carlo simulation of the NA49 detector response. For this simulation a sample of heavy-ion collisions was generated using the VENUS event generator~\cite{Werner:1993uh}. Phase space (represented with transverse momentum and rapidity) was divided into bins in which the efficiency was calculated as the ratio between the number of reconstructed and simulated negatively charged pions:
	\begin{equation}
	\label{efficiency}
	\epsilon = \frac{N^{\pi^-}_{sim}}{N^{\pi^-}_{reco}}.
	\end{equation}
	The resulting efficiency map is presented in Fig.~\ref{Fig:corrections} (right). This map is used to calculate weights in Eq. \ref{Eq:q1_TPC} defined as $w_i = 1 / \epsilon$ for the kinematic bin. 
	
	The total systematic uncertainty is yet to be evaluated. Here first checks related to detection inefficiency, azimuthal non-uniformity and analysis technique are reported.  
	
\begin{figure}[htb]
		\noindent
		\includegraphics[width=0.49\textwidth]{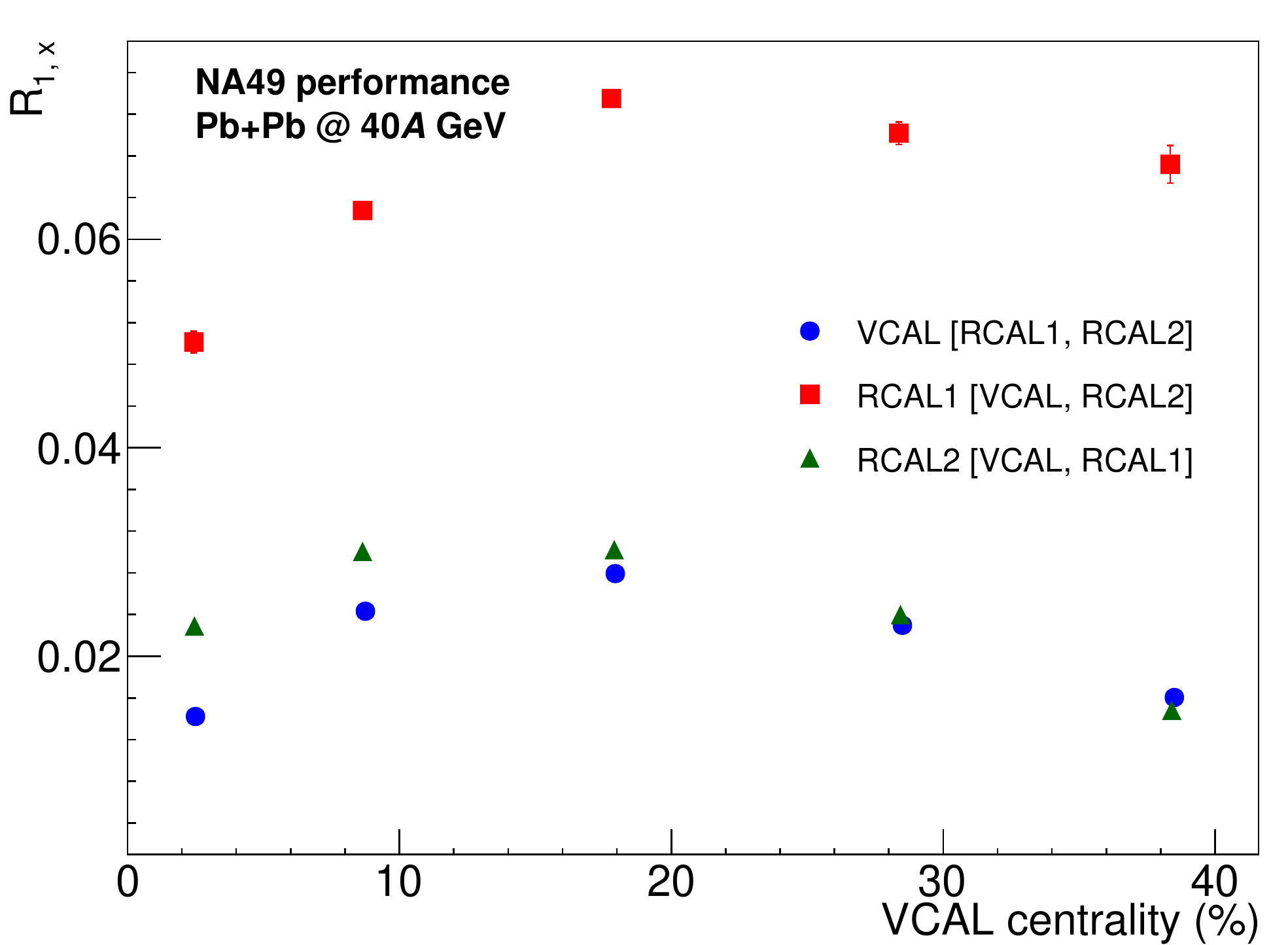}
		\includegraphics[width=0.49\textwidth]{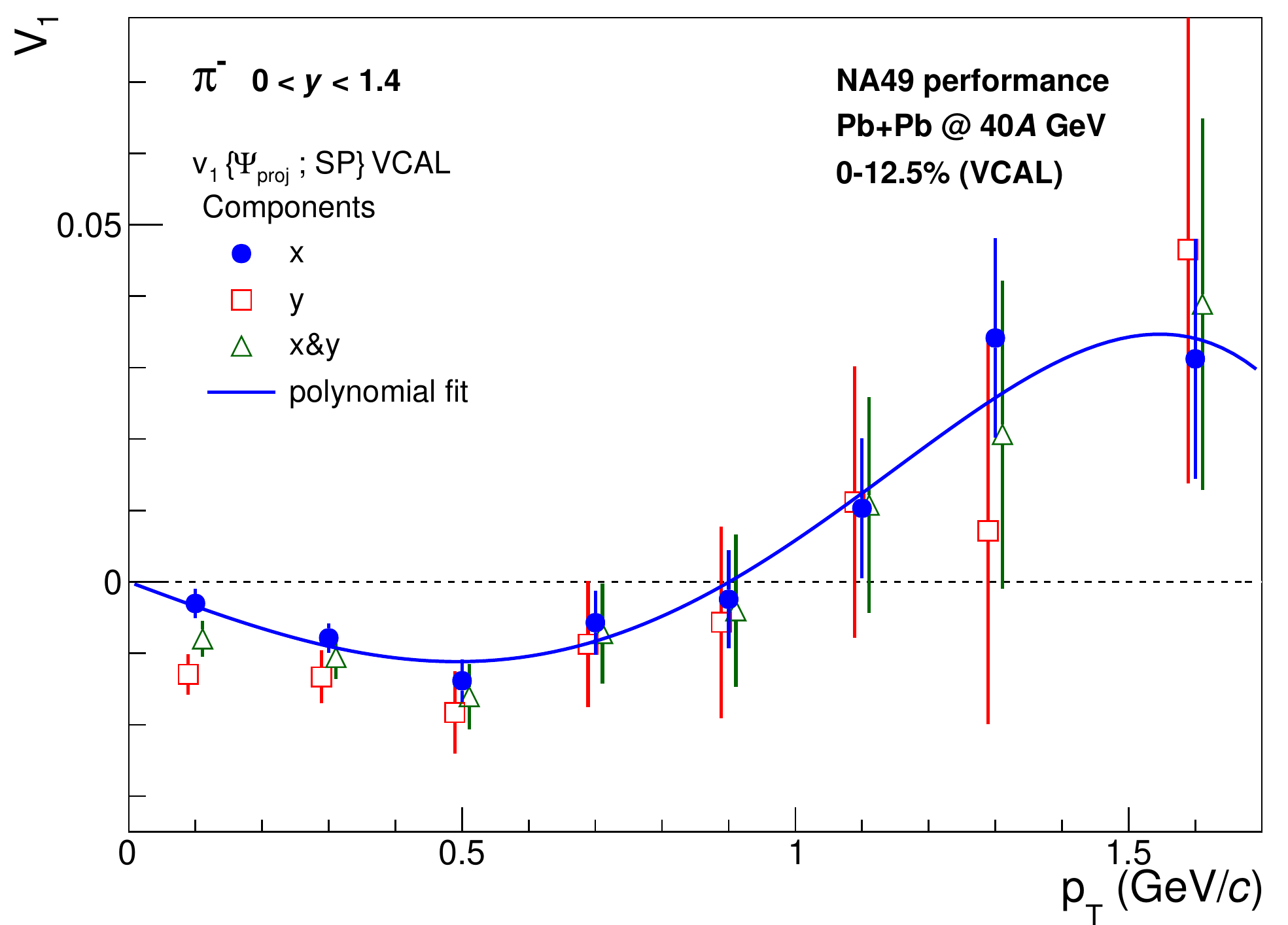}
		\caption{
			Left: Resolution $R_{1, x}$ of spectator symmetry plane angles estimated with VCAL, RCAL1 and RCAL2.
			Right: Directed flow ($v_1$) of negatively charged pions for central collisions measured relative to VCAL using $x$ and $y$ components of \textbf{q} and \textbf{Q}-vectors.}
		\label{Fig:syst}
	\end{figure}

	Due to overlap in pseudorapidity between TPC and RCAL there could be an auto correlation between their \textbf{Q}-vectors, which makes it difficult to do charged hadron flow analysis with the event plane estimated with RCAL.
	Nevertheless, RCAL can be used as a reference detector to extract the symmetry plane resolution of VCAL. 
	Figure~\ref{Fig:syst} (left) shows the resolution $R_{1, x}$ (Eq.~\ref{resolution}) of the spectator symmetry plane angles estimated with components of VCAL, RCAL1 and RCAL2.
	
	Independent $v_1$ estimates with $x$ and $y$ components  of \textbf{q} and \textbf{Q}-vectors were found to be consistent within statistical uncertainties (Fig.~\ref{Fig:syst}, right).
	Owing to the difference in TPC acceptance for $x$ and $y$ directions the correlation for $x$ components are better defined statistically and the presented preliminary results are obtained using $x$ components only.
	
	\section{Results}
	\label{results}
	
	Results are reported for negatively charged pions ($\pi^{-}$) produced by strong interaction process and weak decays (within the TPC acceptance) for 0-12.5\% and 12.5-33.5\% centrality classes. Only statistical uncertainties are shown.
	
	\begin{figure}[htb]
		\noindent
		\includegraphics[width=0.49\textwidth]{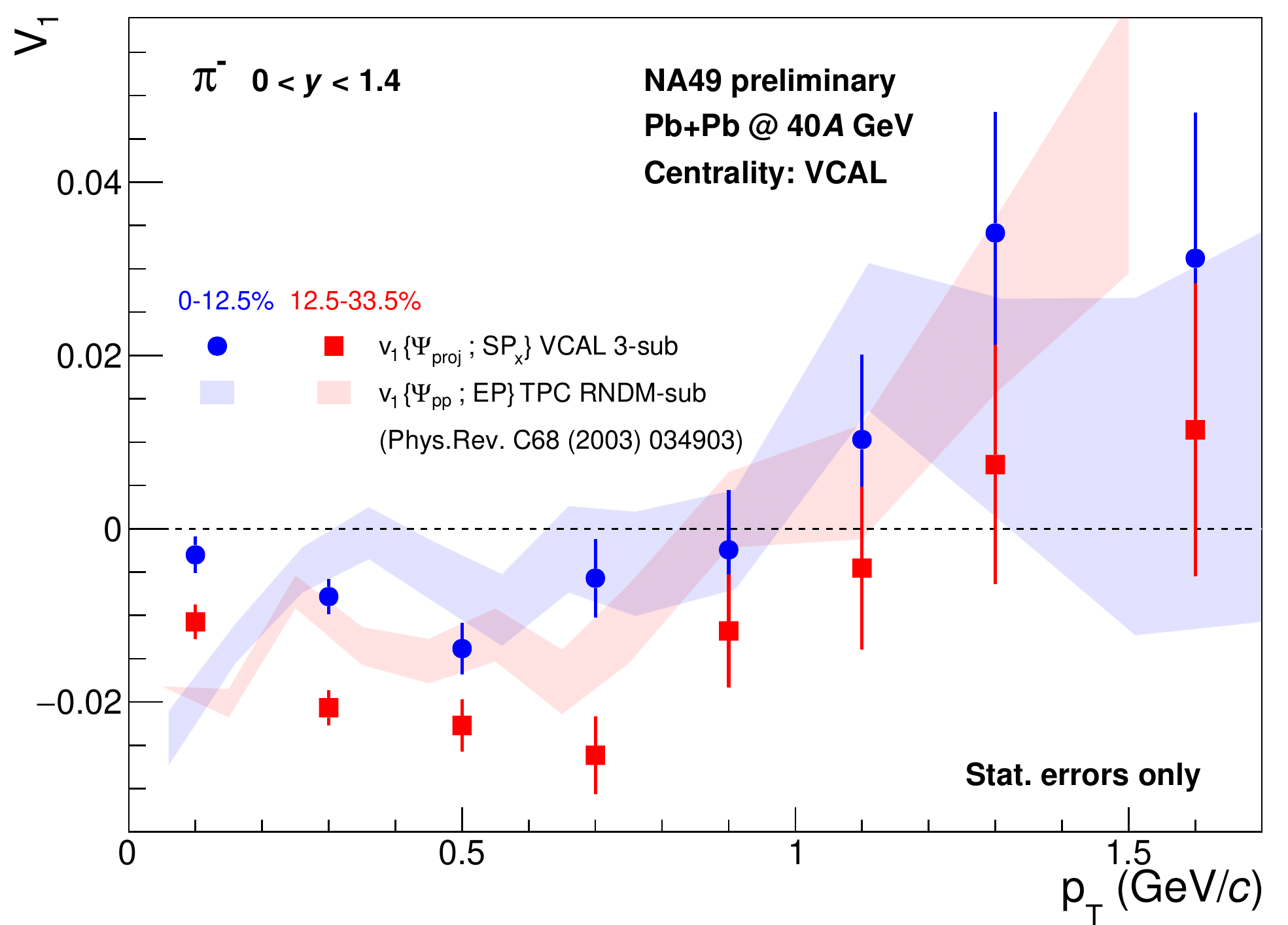}
		\includegraphics[width=0.49\textwidth]{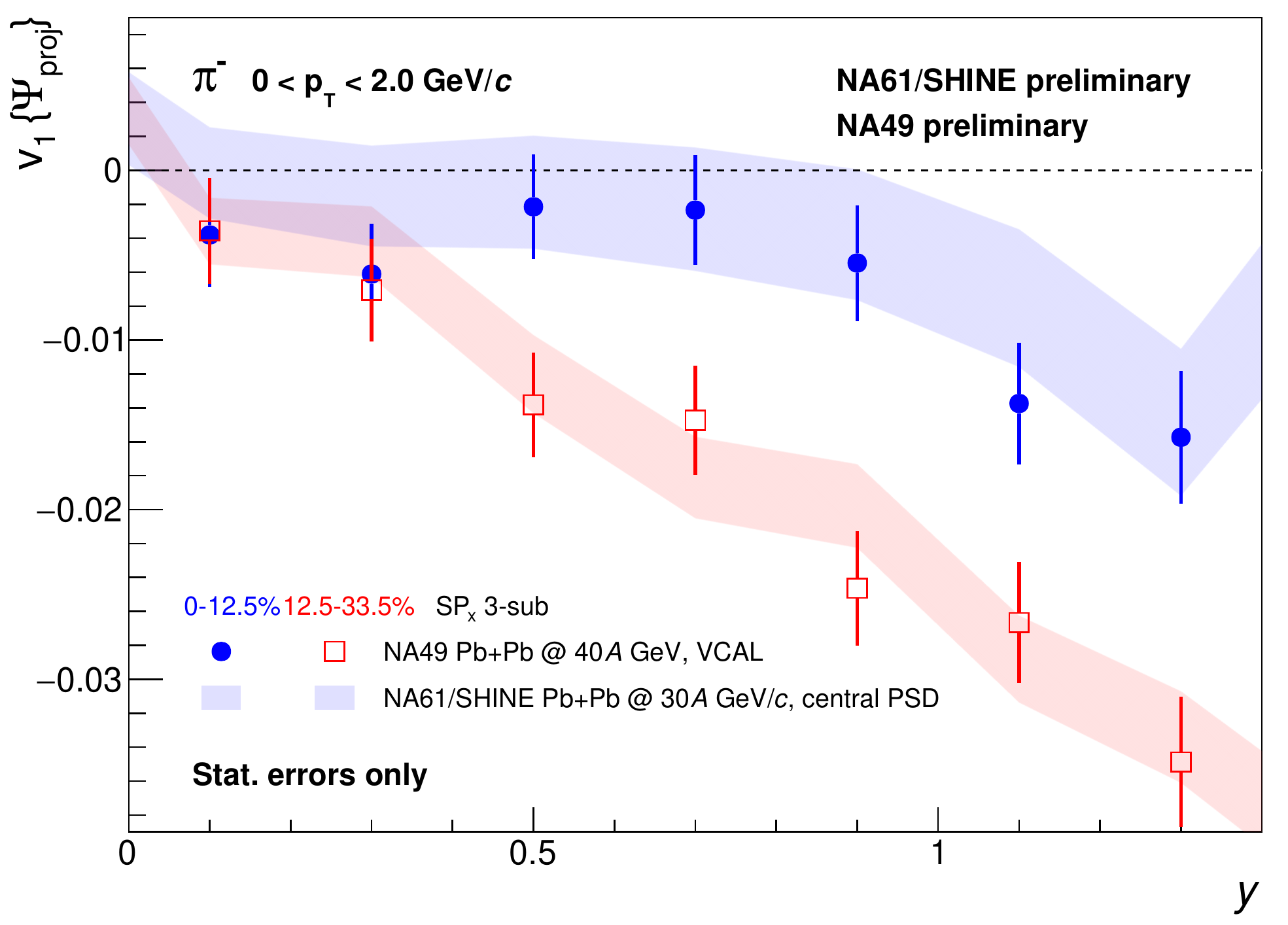}
		\caption{Directed flow ($v_1$) of negatively charged pions in different centrality classes.
			Left: $v_1$ vs. transverse momentum ($p_T$). 
			Right: $v_1$ vs. rapidity ($y$). 
			Only statistical errors are shown.}
		\label{Fig:centrality_comp}
	\end{figure}

	Figure~\ref{Fig:centrality_comp} (left) shows NA49 results for directed flow $v_1$ vs. transverse momentum ($p_T$) measured relative to the projectile spectator symmetry plane in Pb+Pb collisions at 40\AGeV~estimated with VCAL.
	Results are compared to those previously measured by the NA49 Collaboration \cite{Alt:2003ab} relative to the participant plane and corrected for global momentum conservation \cite{Borghini:2002mv}.
	One observes differences between values of $v_1$ relative to the projectile spectator and the participant planes. 
	
	Figure~\ref{Fig:centrality_comp} (right) shows results for $v_1$ vs. rapidity in Pb+Pb collisions at 40\AGeV. 
	The measurements are compared to the new preliminary results \cite{qm_proceedings} for directed flow relative to the spectator plane in Pb+Pb collisions at 30\AGeV/$c$ reported recently by the \SHINE~Collaboration. Despite the small difference in collision energy, good agreement is observed between $v_1$ measured using NA49 and \SHINE~ data.
	
	\section{Summary}
	\label{summary}
	
	Preliminary results for directed flow of negatively charged pions relative to the projectile spectator symmetry plane in Pb+Pb collisions at 40\AGeV~are presented for centrality ranges 0-12.5\% and 12.5-33.5\% as a function of transverse momentum and rapidity.
	Results are corrected for effects of detector azimuthal non-uniformity and detection inefficiency.
	Possible sources of systematic uncertainties are discussed.
	
	Good agreement between directed flow relative to the projectile spectator plane measured by the \SHINE~Collaboration for beam energies of 30\AGeV~ and the NA49 results for 40\AGeV~ is observed.
	Comparison with previously published measurements relative to participant plane shows a dependence of $v_1$ on the symmetry plane estimator.
	
	\section{Acknowledgements}
	\label{acknowledgements}
	
	This work was partially supported by the Ministry of Science and Education of the Russian Federation, grant N 3.3380.2017/4.6, and by the National Research Nuclear University MEPhI in the framework of the Russian Academic Excellence Project (contract No. 02.a03.21.0005, 27.08.2013).
	
	\bibliography{references}
	\bibliographystyle{woc}
	
\end{document}